\documentclass[preprint]{revtex4}

\usepackage{graphicx}
\usepackage{dcolumn}
\usepackage{amsmath}
\usepackage{amsfonts}
\usepackage{mathrsfs}
\usepackage{physics}
\usepackage{mathtools}% http://ctan.org/pkg/mathtools  % For making the Sigma in summation closer to values to its right

\usepackage[T1]{fontenc}
\usepackage[latin1]{inputenc}
\usepackage{booktabs}
\usepackage[font=small,labelfont=bf,tableposition=top]{caption}
\usepackage{rotating}
\usepackage{array,multirow}
\usepackage{float}

\newcommand\cev[1]{\overleftarrow{#1}}

\begin{document}

\title{Conservation Laws for Free Particles Interacting with a Thermal Reservoir in Open Quantum Systems}

\author{M. A. Z. Khan$^{1}$}\email[]{muhammadalzafark@gmail.com}
\author{M. Moodley$^{1}$}\email[]{moodleym2@ukzn.ac.za}
\author{F. Petruccione$^{1}$}\email[]{petruccione@ukzn.ac.za }
%\author{S. Maniscalco$^{2}$}\email[]{smanis@utu.fi }

\affiliation{$^{1}$Quantum Research Group, School of Chemistry and Physics, 
University of KwaZulu--Natal, Private Bag X54001, Durban 4000, South Africa}
%\affiliation{$^{2}$QTF Center of Excellence, Turku Center for Quantum Physics, Department of Physics and Astronomy, University of Turku, FI-20014 Turun Yliopisto, Finland}

\begin{abstract}
We construct Lie point symmetries, a closed-form solution and conservation laws using a non-Noetherian approach for a specific case of the Gorini-Kossakowski-Sudarshan-Lindblad equation that has been recast for the study of non-relativistic free particles in a thermal reservoir environment. Conservation laws are constructed subsequently using the Ibragimov method via a solution to the adjoint form of the equation of motion via its corresponding scalaing symmetry. A general computational framework for obtaining all conserved vectors is exhibited some triplets of conserved quantities are calculated in full.  
\end{abstract}

\maketitle


\section{Introduction}
An ``open quantum system'' is a quantum mechanical system which interacts with its
environment. The environment can be external noise or a quantum environment. The
quantum environment can be bosonic or fermionic in nature; the quantum system is subjected to the effects of the environment. The environment contains many degrees of
freedom - infinite in the case of a reservoir or a thermal bath. Of particular interest is the
collective evolution of the system as an ensemble, not the motion of the individual degrees
of freedom. Thus, the foremost intent is to ascertain the effects of the environment on
the open system.

The theoretical rationalism for studying open quantum systems is essentially divided
into six components - 
\begin{enumerate}
\item All quantum systems are truly open; the random and systematic superimposition
of erroneous signals is unavoidable. The Schrödinger equation, which describes
the dynamics of singleton particles, is simply an approximation and does
not epitomize reality as observed by experiments.
\item It is the foundational fabric and substructure to assess and describe open problems
in quantum theory. For example, the quantum-to-classical transition \cite{Schlosshauer1} and the
quantum measurement problem \cite{Schlosshauer2}.
\item Technological applications of quantum theory. Exploiting - for example, quantum
entanglement - for sensing, metrology, computation, communication and simulation \cite{NielsenChuang}. It has been observed experimentally that these quantum features are
vulnerable to noise and thus, they are quickly consumed. As a consequence, there
is a limited lifetime for which any quantum technology can exist. Therefore, the
environments are modelled in order to understand and minimize noise. 
\item Quantum reservoir engineering and quantum optimal control. To minimize the effects of decoherence induced by the environment and to enhance coherence or
engineer types of decoherence that allow the final states of the dynmaics to have
specific entangled states, which serve as resources of quantum computation. Further,
\begin{itemize}
\item to modify some of the parameters of the environment - for example, the
spectral density in order to transfigure it for quantum specific technological
purposes.
\item to add dissipative sources to the isolated system so as to reach some entangled
or target states that are advantageous resources for quantum technologies.
\end{itemize}
\item Quantum simulations. To simulate quantum systems controllably. These allow
for the addressal of open problems - for example, the analog of the Schrödinger
equation for open quantum systems.
\item Quantum probing. Consider a small quantum system such as an atom confined
in a certain optical potential and consider a many-body system such as an optical
lattice or a complex network. Allowing the two systems to interact and being able
to turn on and turn off the interaction strength perfectly, the small quantum system
can be probed to understand the dynamics of the many-body system because the
decohering dynamics is induced upon the probe. Therefore, the properties of the
many-body system can be mapped onto the probe. By immersing the probe, one
wishes to observe that the global property, perhaps the ground state of the many-body
system, could be detected in the type of induced decoherence on the probe. For example, the Bose-Hubbard model for the transition of a superfluid to a Mott insulator regime \cite{MottRegime1,MottRegime2}.  \\
\end{enumerate}

\noindent This paper is organized in the following manner: \\

In \S II, we present the general equations of motion that govern the dissipative dynamics of the system under consideration. Moreover, a discussion of the mathematical and physical properties of the equation of motion is discussed.

In \S III, we derive the Lie point symmetries of the equation of motion and describe the algebra of these symmetries.

In \S IV, we construct a closed-form solution using a scaling symmetry.

In \S V, we rigorously derive the conservation laws of the underlying equation in the chronological order: Construction of formal Lagrangian, realization of the adjoint-form of the equation of motion via the Euler-Lagrange derivative, acquirement of a group-invariant solution of the adjoint equation and a conscientious statement of the general form of the conservation laws. Subsequently, we calculate three conserved vectors.

In \S VI, we discuss the results furnished in this paper 

In \S VII, we provide a contextual background to ancillary formulae used in this research to construct Lie point symmetries and conservation laws. 

%=================================================================================================================================================

\section{The Model}
The dynamics of the open quantum system is described by the Gorini-Kossakowski-Sudarshan-Lindblad master equation / quantum Liovillian equation of motion \cite{GKSL1,GKSL2,GKSL3}
\begin{equation}
\frac{\partial\rho_{s}(t)}{\partial t}=-\imath\left[\hat{H},\rho_{s}(t)\right]+\sum_{k}\gamma_{k}\left[\hat{L}_{k}\rho_{s}(t)\hat{L}_{k}^{\dagger}-\frac{1}{2}\left\{\hat{L}_{k}^{\dagger}\hat{L}_{k},\rho_{s}(t)\right\}\right], \label{1}
\end{equation}
The equation \eqref{1} is an extension of the idealized Schr\"{o}dinger equation for non-interacting systems to include environmental-interaction. Equation \eqref{1} is not the most general equation and arises because of the Born-Markov approximations and recoupling of the system and the environment. The first term in \eqref{1} descibes the unitary dynamics of the system and the second term descibes the dissipative dynamics that is related to the reciprocity between the system and the the environment.

For a non-relativistic free particle exposed to a thermal reservoir, it has been shown in \cite{ThermalRes1,ThemalRes2,ThermalRes4,ThermalRes3} that Equation \eqref{1} reduces to
\begin{equation}
4\pi mh^{2}\frac{\partial\rho}{\partial t}+\imath h^{3}\left(\frac{\partial^{2}\rho}{\partial y^{2}}-\frac{\partial^{2}\rho}{\partial x^{2}}\right)+32\pi^{3}m^{2}\gamma k_{B}T\left(y-x\right)^{2}\rho=0, \label{2}
\end{equation}
where $h\approx 6.626\times 10^{-34}\;J.s$ is the Planck constant, $\imath=\sqrt{-1}$, $m$ is the particle mass, $\gamma$ is the dissipation quantifier, $k_{B}\approx 1.381\times 10^{-23}\;m^{2}.kg/s^{2}.K$ is the Boltzmann constant, $T$ is the environmental temperature and $\rho=\rho(x,y,t)$ is the positively-defined trace 1 reduced density matrix for the system. Principally, Equation \eqref{2} describes the interaction of an unbound particle having uniform potential with a system of high-temperature coupled harmonic oscillators. Compelling physical features are realized from the study of this model. For example, it has been observed that for a free particle system that is in thermal equilibrium with a reservoir, the system localizes when subjected to specific initial conditions. As a result, quantum fluctuations in the dynamical evolution of the system become damped by temperature variations that are present in the medium; mathematically, this entails that the reduced density matrix approximately becomes non-defective and can therefore be scaled anisotropically. Physically, this implies that information from the free particle dissipates into the environment and consequently explains the quantum decoherence phenomenon. This has the remarkable ramification that the system's evolution state over time traces out a path through a higher-dimensional space which is comparatively commensurate to a continuous-time stochastic process with stationary independent increments, \textit{videlicet} a Wiener process (Brownian motion) \cite{ThermalRes3}.

% non-defective = diagnolizable
% anisotropic scaling = a geometric property of a diagnolizable matrix, also called inhomogenous dilation

\section{Lie Point Symmetries}
We consider Lie point symmetries of the form
\begin{equation}
\Gamma=\xi^{x}(x,y,t,\rho_{s})\frac{\partial}{\partial x}+\xi^{y}(x,y,t,\rho)\frac{\partial}{\partial y}+\xi^{t}(x,y,t,\rho)\frac{\partial}{\partial t}+\eta(x,y,t,\rho)\frac{\partial}{\partial\rho}. \label{3}
\end{equation}
Acting the second-order extension operator \eqref{A.2.1} on \eqref{3} together with the condition 
\begin{equation}
\frac{\partial\rho}{\partial t}=\frac{\imath\hbar^{3}\left(\frac{\partial^{2}\rho}{\partial x^{2}}-\frac{\partial^{2}\rho}{\partial y^{2}}\right)-32\pi^{3}m^{2}\gamma k_{B}T\left(y-x\right)^{2}\rho}{4\pi m\hbar^{2}}. \label{4}
\end{equation}
we obtain the $\left(7+\infty\right)$-dimensional subalgebra of symmetries, spanned by the vector fields
\begin{align}
\Gamma_{1}=&\;\frac{\partial}{\partial t}, \qquad\qquad\qquad\;\;\; \Gamma_{2}=\rho\frac{\partial}{\partial t}, \nonumber \\
\Gamma_{3}=&\;\frac{\partial}{\partial x}+\frac{\partial}{\partial y}, \qquad\qquad \Gamma_{4}=t\frac{\partial}{\partial x}+t\frac{\partial}{\partial y}+\frac{2\pi\imath m}{h}\left(x-y\right)\rho\frac{\partial}{\partial\rho}, \nonumber \\
\Gamma_{5}=&\;\left(\frac{x+2y}{2}\right)\frac{\partial}{\partial x}+\left(\frac{2x+y}{2}\right)\frac{\partial}{\partial y}+t\frac{\partial}{\partial t}-\frac{\rho}{2}\frac{\partial}{\partial\rho}, \nonumber \\
\Gamma_{6}=&\;\left(\frac{4\pi k_{B}Tt^{2}-\imath h}{4\pi k_{B}T}\right)\frac{\partial}{\partial x}+t^{2}\frac{\partial}{\partial y}+\frac{4\pi\imath m}{h}t\left(x-y\right)\rho\frac{\partial}{\partial\rho}, \nonumber \\
\Gamma_{7}=&\;\left(\frac{4\pi k_{B}Tt^{3}-3\imath ht}{4\pi k_{B}T}\right)\frac{\partial}{\partial x}+t^{3}\frac{\partial}{\partial y}+\rho\left[\frac{mhx+6\pi mk_{B}T\imath t^{2}\left(x-y\right)}{hk_{B}T}\right]\frac{\partial}{\partial\rho}, \nonumber \\
\Gamma_{\infty}=&\;\Sigma(x,y,t)\frac{\partial}{\partial\rho}, \label{5}
\end{align}
where $\Sigma(x,y,t)$ is any solution that satisfies \eqref{2}. The algebra of the symmetries \eqref{5} is $\left\{\mathcal{A}_{1}\oplus_{s}\mathbb{W}\right\}\oplus_{s}\mathcal{A}^{a}_{3,5}$, where $\mathbb{W}$ is the Heisenberg-Weyl algebra.

\section{A Group Invariant Solution}
Using $\Gamma_{3}$ from \eqref{5}, we obtain the invariants
\begin{equation}
\mu=y-x,\quad \nu=t, \quad \Phi(\mu,\nu)=\rho(x,y,t). \label{6}
\end{equation}
Substituting \eqref{6} into \eqref{2}, we obtain the reduced equation
\begin{equation}
\hbar^{2}\frac{\partial\Phi}{\partial\nu}+8\pi^{2}m^{2}\gamma k_{B}T\mu^{2}\Phi=0. \label{7}
\end{equation}
Solving \eqref{7} and reverting to the original coordinates in \eqref{6}, we obtain the solution
\begin{equation}
\rho(x,y,t)=\varrho(y-x)\exp\left[-\frac{8\pi^{2}m^{2}\gamma k_{B}T}{\hbar^{2}}\left(y-x\right)^{2}t\right], \label{8}
\end{equation}
where $\varrho(y-x)$ is some arbitrary function of integration. 

%========================================================================================================================================
\section{Conservation Laws}
Using the Ibragimov method \cite{Ibragimov1,Ibragimov2}, we construct the formal Lagrangian for \eqref{2}
\begin{equation}
\mathscr{L}=\vartheta(x,y,t)\left[4\pi h^{2}\frac{\partial\rho}{\partial t}+\imath h^{3}\left(\frac{\partial^{2}\rho}{\partial y^{2}}-\frac{\partial^{2}\rho}{\partial x^{2}}\right)+32\pi^{3}m^{2}\gamma k_{B}T\left(y-x\right)^{2}\rho\right]. \label{9}
\end{equation}
Applying the Euler-Lagrange derivative \eqref{A.5} to \eqref{9}, we obtain the adjoint equation
\begin{equation}
\imath h^{3}\left(\frac{\partial^{2}\vartheta}{\partial y^{2}}-\frac{\partial^{2}\vartheta}{\partial x^{2}}\right)-4\pi h^{2}\frac{\partial\vartheta}{\partial t}+32\pi^{3}m^{2}k_{B}T\left(y-x\right)^{2}\gamma\vartheta=0. \label{10}
\end{equation}
Following the same procedure as in \S III and \S IV, it is easily verified that Equation \eqref{10} admits the Lie point symmetry, \textit{inter alia}, and corresponding invariant solution
\begin{equation}
\cev{\Gamma}_{3}=\frac{\partial}{\partial x}+\frac{\partial}{\partial y}, \qquad\qquad \vartheta(x,y,t)=\varpi(y-x)\exp\left[\frac{8\pi^{2}m^{2}\gamma k_{B}T}{h^{2}}\left(y-x\right)^{2}t\right], \label{11}
\end{equation}
where $\varpi(y-x)$ is some arbitrary function of integration. 

We consider conservation laws of the form $\boldsymbol{\phi}=\left\langle\phi^{x},\phi^{y},\phi^{t}\right\rangle$. From \eqref{A.6.1} and \eqref{A.6.2},  we discern that the general form of the conserved currents are
\begin{align}
\bar{\phi}^{t}=&\;\mathscr{L}\xi^{t}+4\pi h^{2}\left(\eta-\xi^{t}\frac{\partial\rho}{\partial t}-\xi^{x}\frac{\partial\rho}{\partial x}-\xi^{y}\frac{\partial\rho}{\partial y}\right)\vartheta, \tag{12.1}\label{12.1} \\
\bar{\phi}^{x}=&\;\mathscr{L}\xi^{x}+\imath h^{3}\left(\eta-\xi^{t}\frac{\partial\rho}{\partial t}-\xi^{x}\frac{\partial\rho}{\partial x}-\xi^{y}\frac{\partial\rho}{\partial y}\right)\frac{\partial\vartheta}{\partial x}-\imath h^{3}\frac{\partial}{\partial x}\left(\eta-\xi^{t}\frac{\partial\rho}{\partial t}-\xi^{x}\frac{\partial\rho}{\partial x}-\xi^{y}\frac{\partial\rho}{\partial y}\right)\vartheta, \tag{12.2}\label{12.2} \\
\bar{\phi}^{y}=&\;\mathscr{L}\xi^{y}-\imath h^{3}\left(\eta-\xi^{t}\frac{\partial\rho}{\partial t}-\xi^{x}\frac{\partial\rho}{\partial x}-\xi^{y}\frac{\partial\rho}{\partial y}\right)\frac{\partial\vartheta}{\partial y}+\imath h^{3}\frac{\partial}{\partial y}\left(\eta-\xi^{t}\frac{\partial\rho}{\partial t}-\xi^{x}\frac{\partial\rho}{\partial x}-\xi^{y}\frac{\partial\rho}{\partial y}\right)\vartheta. \tag{12.3}\label{12.3}
\end{align}
Using \eqref{12.1}-\eqref{12.3} and the form of the adjoint equation in \eqref{11}, we demonstrate some of the conservation laws below.
\subsection{Conservation of Energy}
\begin{align}
\phi^{t}_{1}=&\;\varpi(y-x)\exp\left[\frac{8\pi^{2}m^{2}\gamma k_{B}T}{h^{2}}\left(y-x\right)^{2}t\right]\left[32k_{B}m^{2}\pi^{3}\gamma T\left(x-y\right)^{2}+\imath h^{2}\left(\frac{\partial^{2}\rho}{\partial y^{2}}-\frac{\partial^{2}\rho}{\partial x^{2}}\right)\right], \tag{13.1}\label{13.1} \\
\phi^{x}_{1}=&\;\imath h\exp\left[\frac{8\pi^{2}m^{2}\gamma k_{B}T}{h^{2}}\left(y-x\right)^{2}t\right]\left\{h^{2}\varpi'(y-x)\frac{\partial\rho}{\partial t}+\varpi(y-x)\left[h^{2}\frac{\partial^{2}\rho}{\partial t\partial x} \right.\right. \nonumber \\
&\left.\left.-16k_{B}m^{2}\pi^{2}T\gamma t\left(x-y\right)\frac{\partial\rho}{\partial t}\right]\right\}, \tag{13.2}\label{13.2} \\
\phi^{y}_{1}=&\;-\imath h\exp\left[\frac{8\pi^{2}m^{2}\gamma k_{B}T}{h^{2}}\left(y-x\right)^{2}t\right]\left\{-h^{2}\varpi'(y-x)\frac{\partial\rho}{\partial t}+\varpi(y-x)\left[h^{2}\frac{\partial^{2}\rho}{\partial t\partial y} \right.\right. \nonumber \\
&\left.\left.\;+16k_{B}m^{2}\pi^{2}T\gamma t\left(x-y\right)\frac{\partial\rho}{\partial t}\right]\right\}. \tag{13.3}\label{13.3}
\end{align}

\subsection{Non-trivial Conservation Law I}
\begin{align}
\phi_{2}^{t}=&\;\rho\varpi(y-x)\exp\left[\frac{8\pi^{2}m^{2}\gamma k_{B}T}{h^{2}}\left(y-x\right)^{2}t\right]\left[32k_{B}m^{2}\pi^{3}\gamma T\left(x-y\right)^{2}\rho+\imath h^{3}\left(\frac{\partial^{2}\rho}{\partial y^{2}}-\frac{\partial^{2}\rho}{\partial x^{2}}\right)\right], \tag{13.4}\label{13.4} \\
\phi_{2}^{x}=&\;\imath h\exp\left[\frac{8\pi^{2}m^{2}\gamma k_{B}T}{h^{2}}\left(y-x\right)^{2}t\right]\left\{h^{2}\rho\varpi'(y-x)\frac{\partial\rho}{\partial t}+\varpi(y-x)\left[h^{2}\frac{\partial\rho}{\partial t}\frac{\partial\rho}{\partial x}\right.\right. \nonumber \\
&\left.\left.\;+h^{2}\rho\frac{\partial^{2}\rho}{\partial t\partial x}-16k_{B}m^{2}\pi^{2}\gamma Tt\left(x-y\right)\rho\frac{\partial\rho}{\partial t}\right]\right\}, \tag{13.5}\label{13.5} \\
\phi_{2}^{y}=&\;-\imath h\exp\left[\frac{8\pi^{2}m^{2}\gamma k_{B}T}{h^{2}}\left(y-x\right)^{2}t\right]\left\{-h^{2}\rho\varpi'(y-x)\frac{\partial\rho}{\partial t}+\varpi(y-x)\left[\frac{\partial\rho}{\partial t}\frac{\partial\rho}{\partial y}\right.\right. \nonumber \\
&\left.\left.\;+h^{2}\rho\frac{\partial^{2}\rho}{\partial t \partial y}+16k_{B}m^{2}\pi^{2}\gamma Tt\left(x-y\right)\rho\frac{\partial\rho}{\partial t}\right]\right\}. \tag{13.6}\label{13.6}
\end{align}

\subsection{Non-trivial Conservation Law II}
\begin{align}
\phi_{3}^{t}=&\;-4\pi h^{2}\exp\left[\frac{8\pi^{2}m^{2}\gamma k_{B}T}{h^{2}}\left(y-x\right)^{2}t\right]\varpi(y-x)\left(\frac{\partial\rho}{\partial x}+\frac{\partial\rho}{\partial y}\right), \tag{13.7}\label{13.7} \\
\phi_{3}^{x}=&\;\exp\left[\frac{8\pi^{2}m^{2}\gamma k_{B}T}{h^{2}}\left(y-x\right)^{2}t\right]\left\{\imath h\left[h^{2}\varpi'(y-x)-16k_{B}m^{2}\pi^{2}\gamma Tt\left(y-x\right)\varpi(y-x)\right]\right. \nonumber \\
&\left.\;\times\left(\frac{\partial\rho}{\partial x}+\frac{\partial\rho}{\partial y}\right)+\varpi(y-x)\left[32k_{B}m^{2}\pi^{3}\gamma T\left(x-y\right)^{2}\rho+4\pi h^{2}\frac{\partial\rho}{\partial t}+\imath h^{3}\left(\frac{\partial^{2}\rho}{\partial y^{2}}-\frac{\partial^{2}\rho}{\partial x^{2}}\right)\right] \right. \nonumber \\
&\left.\;+\imath h^{3}\left(\frac{\partial^{2}\rho}{\partial x\partial y}+\frac{\partial^{2}\rho}{\partial x^{2}}\right) \right\}, \tag{13.8}\label{13.8} \\
\phi^{y}_{3}=&\;\exp\left[\frac{8\pi^{2}m^{2}\gamma k_{B}T}{h^{2}}\left(y-x\right)^{2}t\right]\left\{\imath h\left[h^{2}\varpi'(y-x)-16k_{B}m^{2}\pi^{2}\gamma Tt\left(y-x\right)\varpi(y-x)\right]\right. \nonumber \\
&\left.\;\times\left(\frac{\partial\rho}{\partial x}+\frac{\partial\rho}{\partial y}\right)-\imath h^{3}\varpi(y-x)\left(\frac{\partial^{2}\rho}{\partial x\partial y}+\frac{\partial^{2}\rho}{\partial y^{2}}\right)+\varpi(y-x)\left[32k_{B}m^{2}\pi^{3}\gamma T\left(x-y\right)^{2}\rho\right.\right.\nonumber \\
&\left.\left.\;+4\pi h^{2}\frac{\partial\rho}{\partial t}+\imath h^{3}\left(\frac{\partial^{2}\rho}{\partial y^{2}}-\frac{\partial^{2}\rho}{\partial x^{2}}\right)\right]\right\}. \tag{13.9}\label{13.9}
\end{align}

%==================================================================================================
\section{Discussion}\label{discussion}
Using the Lie point symmetry method, we obtained a $\left(7+\infty\right)$-dimensional set of point symmetries. By examination of the nonzero commutator actions, it was shown that, \textit{inter alia}, a Heisenberg-Weyl algebra was obtained. Upon investigation of the literature, Equation \eqref{2} shares this ubiquitous feature with the Schr\"{o}dinger equations for a homogeneous and harmonic potential energies. Using the scaling symmetry, an invariant solution was constructed in terms of a generic integration function, which serves as the normalization constant.   

Conservation laws were constructed using a non-Noetherian approach. This has the added benefit that one need not derive the equation of motion from Euler-Lagrange equations and guarantees that every symmetry leads to a conservation law; properties that Noether's approach lacks. Moreover, we have provided the general form of all subsequent conservation laws that one wishes to construct at one's own liberty and further demonstrated the construction of three such pairs. Interestingly, the conservation of energy law in \eqref{13.1}-\eqref{13.3} is particularly provocative as one would think that the only energy of the system that is conserved is the Hamiltonian operator. In principal, one would be able to generate a large class, possibly infinite, of conserved energies, contingent upon the choice of the normalization constant so as to ensure its convergence. Finally, we note that equation \eqref{2} is linear and therefore possesses any infinite number of point symmetries and conservation laws; this advocates the appearance of $\Gamma_{\infty}$. The formulae \eqref{12.1}-\eqref{12.3} that we have proposed gives an algorithmic and computational approach to obtaining these conservation laws. 

%==================================================================================================
\section{Appendix}
In this section, we provide a cursory overview of the Lie symmetry method and the Ibragimov method. Following the notation of Olver \cite{Olver}, consider the $n^{\text{th}}$-order PDE of the form
\begin{equation}
\Delta\left(x,y,t,\rho,\frac{\partial\rho}{\partial x},\frac{\partial\rho}{\partial y},\frac{\partial\rho}{\partial t},\frac{\partial^{2}\rho}{\partial x^{2}},\frac{\partial^{2}\rho}{\partial y^{2}},\ldots\right)=0, \tag{A.1}\label{A.1}
\end{equation}
where $\rho=\rho(x,y,t)$. Equation \eqref{A.1} admits Lie point symmetries of the form \eqref{2} with the $n^{\text{th}}$-order extension operator given by
\begin{align}
\text{pr}^{\left[2\right]}\Gamma\equiv \Gamma^{\left[2\right]}=&\;\Gamma+\zeta^{t}\frac{\partial}{\partial\rho_{t}}+\zeta^{xx}\frac{\partial}{\partial\rho_{xx}}+\zeta^{yy}\frac{\partial}{\partial\rho_{yy}}, \tag{A.2.1}\label{A.2.1} \\
\zeta^{x}=&\;D_{x}\left(\eta\right)-D_{x}\left(\xi^{x}\right)\rho_{x}-D_{x}\left(\xi^{y}\right)\rho_{y}-D_{x}\left(\xi^{t}\right)\rho_{t}, \tag{A.2.2}\label{A.2.2} \\
\zeta^{y}=&\;D_{y}\left(\eta\right)-D_{y}\left(\xi^{x}\right)\rho_{x}-D_{y}\left(\xi^{y}\right)\rho_{y}-D_{y}\left(\xi^{t}\right)\rho_{t}, \tag{A.2.3}\label{A.2.3} \\
\zeta^{t}=&\;D_{t}\left(\eta\right)-D_{t}\left(\xi^{x}\right)\rho_{x}-D_{t}\left(\xi^{y}\right)\rho_{y}-D_{t}\left(\xi^{t}\right)\rho_{t}, \tag{A.2.4}\label{A.2.4} \\
\zeta^{xx}=&\;D_{x}\left(\zeta^{x}\right)-D_{x}\left(\xi^{x}\right)\rho_{xx}-D_{x}\left(\xi^{y}\right)\rho_{yx}-D_{x}\left(\zeta^{t}\right)\rho_{tx}, \tag{A.2.5}\label{A.2.5} \\
\zeta^{yy}=&\;D_{y}\left(\zeta^{y}\right)-D_{y}\left(\xi^{x}\right)\rho_{xy}-D_{y}\left(\xi^{y}\right)\rho_{yy}-D_{y}\left(\zeta^{t}\right)\rho_{ty}, \tag{A.2.6}\label{A.2.6} 
\end{align}
and $D_{i}$, for $i\in\left\{x,y,t\right\}$, is the total differential operator. The application of \eqref{A.2.1} on \eqref{A.1} yields a system of linear PDEs
\begin{equation}
\xi^{x}\frac{\partial\Delta}{\partial x}+\xi^{y}\frac{\partial\Delta}{\partial y}+\xi^{t}\frac{\partial\Delta}{\partial t}+\zeta^{t}\frac{\partial\Delta}{\partial\rho_{t}}+\zeta^{xx}\frac{\partial\Delta}{\partial\rho_{xx}}+\zeta^{yy}\frac{\partial\Delta}{\partial\rho_{yy}}=0. \tag{A.3}\label{A.3}
\end{equation}
Equating the various powers of the variable $\Delta$ in \eqref{A.3} to zero produces a set of linear PDEs. Solving the system of PDEs provides functional forms for $\xi^{x}, \xi^{y},\xi^{t}$ and $\eta$ which gives each Lie point symmetry.

The formal Lagrangian is given by
\begin{equation}
\mathcal{L}=\sum\limits_{i=1}^{m}v_{i}F_{i}\left(\mathbf{x},\boldsymbol{u_{\left(i\right)}}\right), \tag{A.4}\label{A.4}
\end{equation}
where $\mathbf{x}=\left(x_{1},x_{2},\ldots,x_{m}\right)$ and $\boldsymbol{u_{\left(i\right)}}=\left(u_{\left(1\right)},u_{\left(2\right)},\dots,u_{\left(s\right)}\right)$. The Euler-Lagrange derivative is given by
\begin{equation}
\frac{\delta}{\delta u^{\alpha}}=\frac{\partial}{\partial u^{\alpha}}+\sum_{r\geq 1}D_{i_{1}}, D_{i_{2}}, \ldots, D_{i_{r}}\frac{\partial}{\partial u^{\alpha}_{i_{1}i_{2}\ldots i_{r}}}, \qquad \alpha=1,2\ldots,m. \tag{A.5}\label{A.5}
\end{equation}
Applying \eqref{A.5} on \eqref{A.4} yields the adjoint form of the underlying equation. Thereafter, one applies the Lie method in order to obtain a set of new symmetries and corresponding invariant solutions of the adjoint equation. Subsequently, one substitutes in the symmetries of the underlying equations together with the invariant solutions of the adjoint equation into
\begin{align}
\varphi^{i}=&\;\mathcal{L}\xi^{i}+W^{\alpha}\left[\frac{\partial\mathcal{L}}{\partial u_{i}^{\alpha}}-D_{j}\left(\frac{\partial\mathcal{L}}{\partial u^{\alpha}_{ij}}\right)+D_{j}D_{k}\left(\frac{\partial\mathcal{L}}{\partial u^{\alpha}_{ijk}}\right)-\ldots\right] \nonumber \\
&\;+D_{j}\left(W^{\alpha}\right)\left[\frac{\partial\mathcal{L}}{\partial u^{\alpha}_{ij}}-D_{k}\left(\frac{\partial\mathcal{L}}{\partial u^{\alpha}_{ijk}}\right)+\ldots\right] \nonumber \\
&\;+D_{j}D_{k}\left(W^{\alpha}\right)\left(\frac{\partial\mathcal{L}}{\partial u^{\alpha}_{ijk}}\right)+\ldots, \tag{A.6.1}\label{A.6.1} \\
W^{\alpha}=&\;\eta^{\alpha}-\sum_{\mathclap{i\in\mathbf{x}=\left(x_{1},x_{2},\ldots,x_{m}\right)}}\xi^{i}u_{i}^{\alpha}, \tag{A.6.2}\label{A.6.2}
\end{align}
in order to obtain the conservation laws of the form
\begin{equation}
\sum_{\mathclap{i\in\mathbf{x}=\left(x_{1},x_{2},\ldots,x_{m}\right)}}D_{i}\varphi^{i}=0. \tag{A.6.3}\label{A.6.3}
\end{equation}

%=================================================================================================
\begin{acknowledgements}
MAZK, MM and FP thank the  University of KwaZulu-Natal of the Republic of South Africa for making this research possible. FP acknowledges that this research is supported by the South African Research Chair Initiative of the Department of Science and Innovation and the National Research Foundation (NRF). 
%SM thanks the University of Turku of Finland for making this research possible. SM acknowledges financial support from the Academy of Finland via the Centre of Excellence (CoE) program.
\end{acknowledgements}

%==================================================================================================

\end{document}